\documentclass[utf8,aps,twocolumn,pre]{revtex4-1}

\usepackage{graphicx}
\usepackage{epstopdf}
\usepackage{listings}	
\usepackage{amssymb}
\usepackage{amsmath}
\usepackage{amsthm}
\usepackage[extension=xxx]{hyperref}
\usepackage{color}
\usepackage[utf8]{inputenc}
\usepackage{lipsum}
\usepackage{breqn}

\def\be{\begin{equation}}
\def\ee{\end{equation}}
\def\bea{\begin{eqnarray}}
\def\eea{\end{eqnarray}}

\begin{document}

\author{Serge Dmitrieff}
 \email{serge.dmitrieff@ijm.fr}
 \author{Nicolas Minc}
\affiliation{
Institut Jacques Monod CNRS UMR7592 and Universite Paris Diderot \\ 75205 Paris Cedex 13 France
}%

\title{Scaling properties of centering forces.}
\date{\today}

\begin{abstract}
Motivated by the centering of biological objects in large cells, we study the generic properties of centering forces inside a ball (or a volume of spherical topology) in $n$ dimensions. We consider two scenarios : autonomous centering (in which distance information is integrated from the agent perspective) and non-autonomous centering (in which distance to the surface is integrated over the whole surface). We find relations between the net centering force and the mean distance$^p$ to the surface. This allows us to find simple scaling laws between the centering force and the distance to the center, as a function of the dimensionality $n$. Interestingly, if the interactions between the agent and the surface are hyper-elastic, the net centering force can still be sub-elastic in the case of autonomous centering. These scaling laws are increasingly violated as the space becomes less convex. Generically, neither scenarios exactly converge to the center of mass of the space.
\end{abstract}

\maketitle

In animal eggs after fertilization, the male pronucleus reaches the center of the cell, seemingly following only geometrical cues : in deformed cells, the pronucleus seems to stop at the center of mass \cite{minc2011influence}. It is remarkable that a small biological object can robustly find the center of the containing space autonomously. It has thus gained a lot of experimental and theoretical attention. The pronucleus creates an aster, a radial structure of stiff elastic filaments called microtubules. Motors in the egg volume and at the surface pull on microtubules, creating pulling  force on microtubules, that allows efficient aster centration \cite{tanimoto2016shape,tanimoto2018physical,grill2005spindle,wuehr2009does}. These forces should scale in length$^p$, were $p$ is an exponent depending on the details on the pulling mechanisms and the availabitlity of motors \cite{kimura2011novel,minc2011influence}. Pushing forces from microtubules on the egg surface have also been proposed to favor centering, but experimental evidence in many species highlight the dominant role of pulling forces \cite{wuehr2009does}. Numerical simulations studied the centering of the aster, taking into account pusing and pulling \cite{zhu2010finding}, as well as the dynamics and flexibility of the microtubules \cite{letort2016centrosome}. However, the generic properties of this centering mechanism were not studied. In particular, how the forces depend on the egg shape, and on the force exponent $p$, remains unknown. 

This problem can be seen in the more general context of an agent needing to find a location autonomously, i.e. without relying on external directions. This is of special interest for the localization of autonomous drones, in the absence of GPS signals. In this article we will investigate how an agent can find the center of a space by integrating an information being the distance to the surface. We will compare it to non-autonomous centering.

\begin{figure}[b]
	\includegraphics[width=\linewidth]{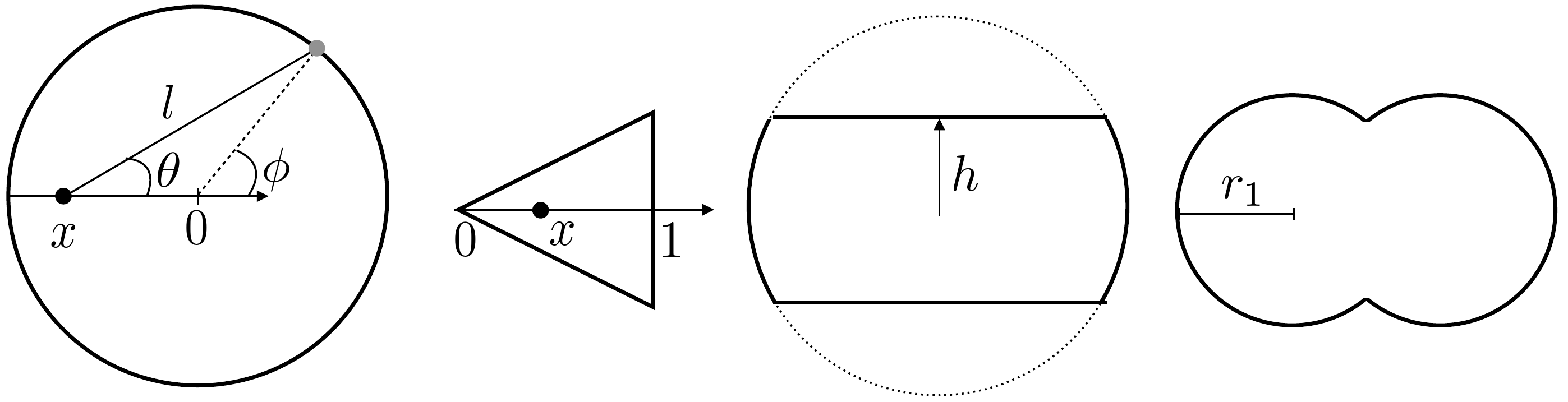} 
	\caption{ \label{illus_shapes}
		Illustration of the shapes considered, in the $Oxy$ plane. From left to right : ball, cone, truncated ball, doublet.
	}
\end{figure}

\paragraph*{Centering in n-balls} We first considered an agent at a distance $x$ from the center, in a $n$-dimensional ball of radius $1$, see Fig. \ref{illus_shapes}. We will consider either  {\em autonomous} centering, in which the agent integrates information over all angles $\theta$ around it, or {\em non-autonomous} centering, in which the information is integrated over the surface. Lets us first consider autonomous centering ; we can call $l(x,\theta)$ the distance between the agent and the point on the surface situated at an angle $\theta$ in the plane $Ox,\mathbf{u}_\theta$, with $\mathbf{u}_\theta$ the direction vector for angle $\theta$. We define $\bar{l}_n^p(x)$ the mean value of $l(x,\theta)^p$ averaged over all $\theta$  (the weight of each angle $\theta$ depending on $\theta$ and $n$), see Eq. \ref{def_dist} . For simplicity, we will assume the space to be symmetric around the axis $Ox$.

We could consider two centering mechanisms of autonomous centering : $(i)$ maximizing $\bar{l}_n^p(x)$, and $(ii)$ pulling towards the surface with a projected force $f_n^p(x,\theta)=l(x,\theta)^p \mathbf{u}_\theta \mathbf{.} \mathbf{u}_x$, averaged over all $\theta$. The later is the strategy adopted by the male pro-nucleus to find the center of the egg. The former yields a force $g_n^p(x)=\partial_x \bar{l}_n^p(x)$ ; an agent implementing a such strategy thus needs a memory in order to compute the gradient, while the pulling strategy $(ii)$ requires no memory. We will see that both methods are highly related and we will focus on $(ii)$, pulling mediated centering. Note that we will only discuss $p>0$ as centering with $p<0$ requires pushing forces, rather than pulling, to achieve centering ; the behaviour of $p<0$ will be mentioned in the discussion.



When considering {\em autonomous centering}, the mean distance$^p$ to the surface, from the agent's perspective, is  :
\begin{eqnarray} \label{def_dist}
\bar{l}_n^p(x)= \frac{1}{ \alpha_n } \int_0^\pi \sin^{n-2}{(\theta)} l(x,\theta)^p d\theta \, , \\
\alpha_n = \frac{\Gamma[\frac{n-1}{2}]}{\Gamma[\frac{n}{2}]} \sqrt{\pi} \, , \\
l(x,\theta)=-x \cos{\theta} + \sqrt{1 - x^2 \sin{\theta} }  \, , \label{def_l_sphere}
\end{eqnarray}
in which $\Gamma$ is the gamma function. For non-spherical spaces, only Eq. \ref{def_l_sphere} needs to be altered. The net centering force by pulling is :
\begin{eqnarray}
\bar{f}_n^p(x)= \frac{-1}{\alpha_n } \int_0^\pi \cos{\theta} \sin^{n-2}{(\theta)} l(x,\theta)^p d\theta \, . \label{def_force}
\end{eqnarray}

\begin{figure}[t]
	\includegraphics[width=\linewidth]{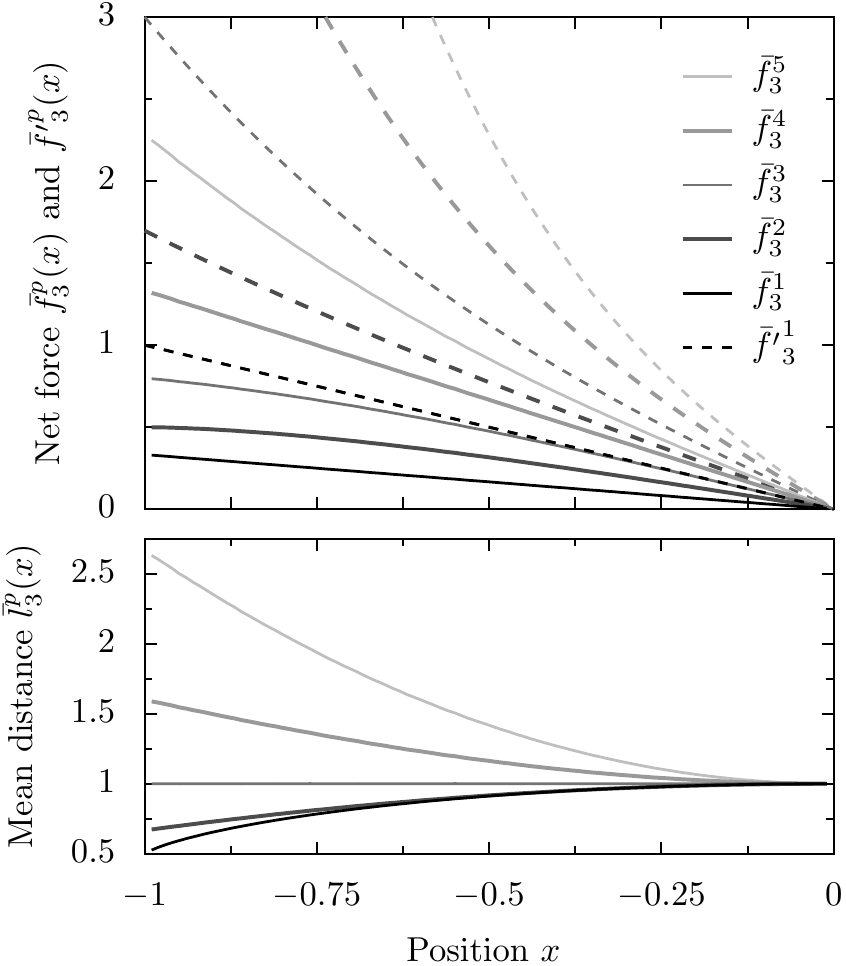} 
	\caption{ \label{fig_1_3D}
		{\em Top} : Net autonomous centering force $\bar{f}_3^p(x)$ (solid lines) and non-autonomous  $\bar{f'}_3^p$ (dashed lines) as a function of $x$ for various values of $p$ in a 3-ball. $\bar{f}_3^p$ is hypo-elastic for $1<p<4$, elastic for $p=1,3$ and hyper-elastic for $p>4$. {\em Bottom} : the mean distance to the surface $\bar{l}_3^p(x)$ is a decreasing function of $x$ for $0<p<3$, constant for $p=3$, and increasing for $p>3$. $\bar{f}_n^p(x)$ and  $\bar{l}_n^p(x)$ were computed numerically by integrating $10^4$ angular elements.
	}
\end{figure}

Unfortunately, we could not solve $\bar{l}_n^p$ and $\bar{f}_n^p$ analytically. However, it is possible to show that : 
\begin{eqnarray}
g_n^p(x) = p \left(\frac{n-p}{p-1} \right) \bar{f}_n^{p-1}(x)  \, . \label{forces_np}
\end{eqnarray}

An important result, is that $g_n^p$ is a centering force for $p<n$ and promotes decentering for $p>n$. Moreover, $\forall n$, $g_n^n(x)=0$. Indeed, $\bar{l}_n^n$ is a measure of the volume visible from the point $x$ (up to a constant prefactor) and should thus not depend upon $x$ if the space is convex \cite{beer1973index}. Because of the normalization by $\alpha_n$, we have $\bar{l}_n^n=1$ if the space is spherical and of radius $1$.

We can now use the fact that for any convex space, 
\begin{eqnarray}
\partial_x l(x,\theta) = - \cos \theta - \frac{1}{l(x,\theta)} \sin \theta \partial_\theta l(x,\theta) 
\end{eqnarray}
And  we find that, for any convex shape : 
\begin{eqnarray} \label{f_np_of_l_np}
\partial_x \bar{f}_n^p= \frac{\alpha_{n+2}}{\alpha_n} p \left( \frac{p-n-1}{p-1} \right)  \bar{l}_{n+2}^{p-1}  - p \left(  \frac{p-n}{p-1} \right) \bar{l}_n^{p-1} . \,
\end{eqnarray}
Using $\bar{l}_n^n=1$, we find :
\begin{eqnarray}
\bar{f}_n^{n+1}(x)=-\bar{l}_n^n \frac{n+1}{n}x \, .
\end{eqnarray}




Therefore, in $n$ dimensions, the net ($n+1$)-force $\bar{f}_n^{n+1}(x)$ is linear with $x$, i.e. the agent centers elastically inside any convex shape. Note that $\bar{f}_n^1$ is also linear in $x$, see equation \ref{f_np_of_l_np}. We thus wondered what happened to $\bar{f}_n^p$ for $1 < p < n+1$ and $p>n+1$. Because of equation \ref{forces_np}, this will also yield the behavior of $g_n^p$.

Before going any further, we can highlight the difference with {\em non-autonomous centering}, in which the points on the surface pull on the agent. This is a generalization of Newton's shell for forces of different exponents, and for any dimension. The net projected pulling force $\bar{f'}_n^p(x)$ can be written :
\begin{eqnarray} \label{def_newt_force}
\bar{f'}_n^p(x) = \frac{-1}{\alpha_n} \int_0^\pi \left(\frac{\cos{\phi}-x}{l'(x,\phi)} \right) \sin^{n-2}{(\phi)} l'(x,\phi)^p d\phi \, ,\quad \\
l'(x,\phi)=\sqrt{1+x^2-2 x \cos{\phi} } \, . \quad  \quad \label{def_l_sphere_Newton}
\end{eqnarray}
In a sphere, it is possible to solve $\bar{f'}_n^p(x)$ analytically, to find, with $_2F_1$ the $(2,1)$ hypergeometric function :
\begin{dmath}
	\bar{f'}_n^p(x)=-(1-x)^{p-1} \left( _2F_1\left(\frac{n+1}{2},\frac{1-p}{2};n;-\frac{4 x}{(x-1)^2}\right)
	+(x-1) _2F_1\left(\frac{n-1}{2},\frac{1-p}{2};n-1;-\frac{4 x}{(x-1)^2}\right)\right)
\end{dmath}
Note that $\bar{f'}_3^0(x)=-2x/3$ is the net force of cortex-mediated centering that has been proposed for centering in some eggs \cite{grill2005spindle}.

While we could not solve analytically $\bar{f}_n^p(x)$ for any $n,p$, we could integrate it numerically by discretizing equation \ref{def_force}. In 3D, we find that  $\bar{l}_3^p$ is constant for $p=0,3$ ; it is minimum at $x=0$ for $p>3$ and otherwise maximum, Fig. \ref{fig_1_3D}, bottom. $\bar{f}_3^p$ is linear for $p=1,4$, sub-linear in between, and superlinear for $p>4$, but always promotes centering, Fig. \ref{fig_1_3D}, top. It is quite remarkable that a sum of hyperelastic forces on the agent results in a net hypoelastic behaviour. This is not the case for a Newton-shell type centering force, in which centering is hyperelastic when $p>1$,  Fig. \ref{fig_1_3D}, top. 

\begin{figure}[t]
	\includegraphics[width=\linewidth]{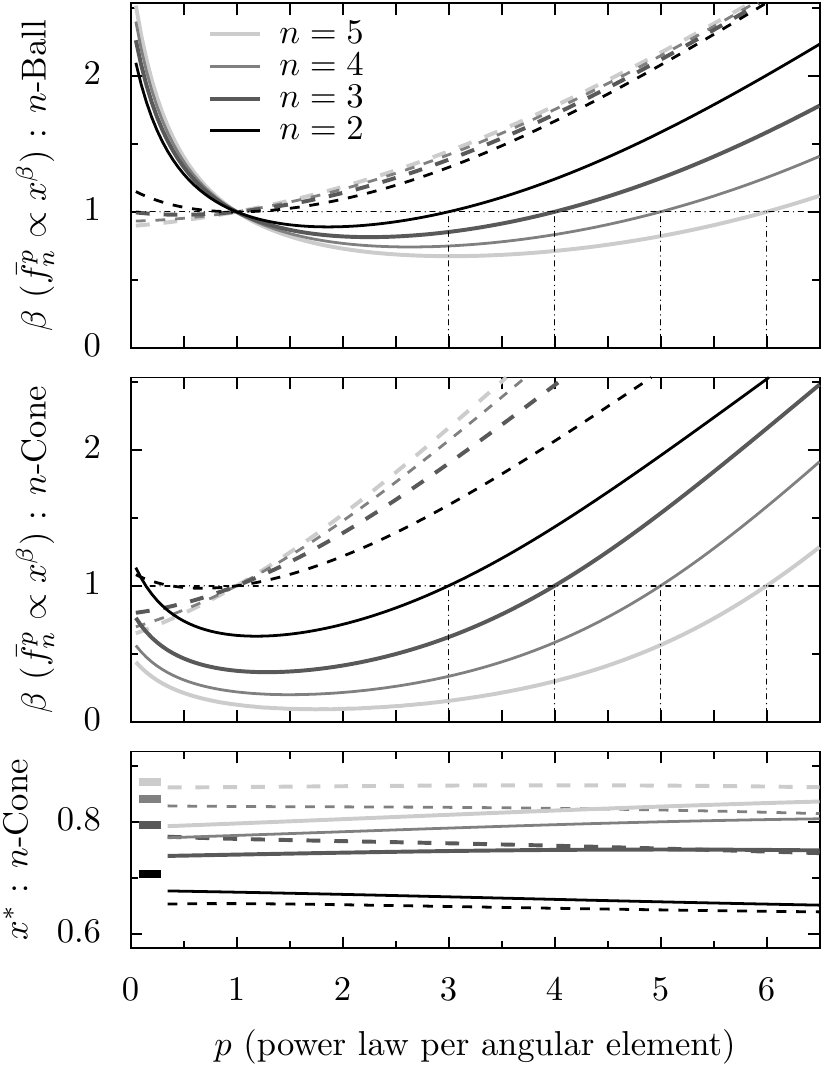} 
	\caption{\label{fig_pwlaw}
		{\em Top} : power law $\beta$, fitting $\bar{f}_n^p(x) \propto x^\beta$ (autonomous centering, solid lines) and  $\bar{f'}_n^p(x) \propto x^\beta$ (surface-mediated centering, dashed lines) in a $n$-ball for different $n$. {\em Middle} : same as above, for centering in a $n$-cone. {\em Bottom} : equilibrium position $x^*$ such that $\bar{f}_n^p(x^*)=0$ (solid lines) and such that $\bar{f'}_n^p(x^*)=0$ (dashed lines) for several $n$. Thick markers on the left indicate the center of mass of the volume.
	}
\end{figure}

We then fitted  $\bar{f}_n^p(x)$ and $\bar{f'}_n^p(x)$, by a power law $x^\beta$ which allowed us to generalize the $3D$ results : the net pulling force  $\bar{f}_n^p(x)$ is elastic for $p=1$ and $p=n+1$, hypo-elastic in between, and hyper-elastic for $p<1$ and $p>n+1$, Fig. \ref{fig_pwlaw}, top. Surface-mediated centering is systematically hyperelastic for $p>1$, with a scaling depending little on $n$, Fig. \ref{fig_pwlaw}.

\paragraph*{Centering in non-spherical spaces}
As mentioned, the space-invariance of $\bar{l}_n^n$ comes from it being a measure of the visible volume, and the relation $\bar{f}_n^{n+1}(x)\propto x$, should hold for any convex space. We thus verified these relationships in the case of a (hyper)cone symmetrical around $Ox$ (see Fig. \ref{illus_shapes}), by integrating numerically Eqns. \ref{def_dist},\ref{def_force}, \ref{def_newt_force}, with a redefinition of the lengths in equations \ref{def_l_sphere},\ref{def_l_sphere_Newton} to match the conical shape. The scaling law  $\bar{f}_n^{n+1}(x)\propto x$ and $\bar{f'}_n^{1}(x)\propto x$   were still valid, Fig. \ref{fig_pwlaw}, middle. Moreover, $\bar{f}_n^{p}(x)$ was under-linear for $1<p<n+1$ and super-linear for $p>n+1$ as in a $n$-ball. However, $\bar{f}_n^{1}(x)$ was not linear with $x$ for $n>1$. Thus, the scaling results we found for $\bar{f}_n^{n+1}$ and $\bar{f'}_n^1$ indeed hold, while $\bar{f}_n^{1}\propto x$ seems valid only for the $n$-ball. 

As mentioned, in non-spherical cells, the aster seems to find the center of mass \cite{minc2011influence}. We therefore inquired the equilibrium position $x^*$ of the various centering mechanisms. Surprisingly, neither mechanism converged strictly to the center of mass of the space, Fig.\ref{fig_pwlaw}, bottom. Interestingly, $x^*$ was not even necessarily monotonous in $p$ (e.g. for $n=3$). This highlights the non-trivial properties of centering forces. While this finding is theoretically interesting, it remains to be seen whether the difference is within the range of experimental resolution in the case of pronucleus centering. More generally, this is to be kept in mind when empirically determining the center of a space by such methods.

\begin{figure}[b]
	\includegraphics[width=\linewidth]{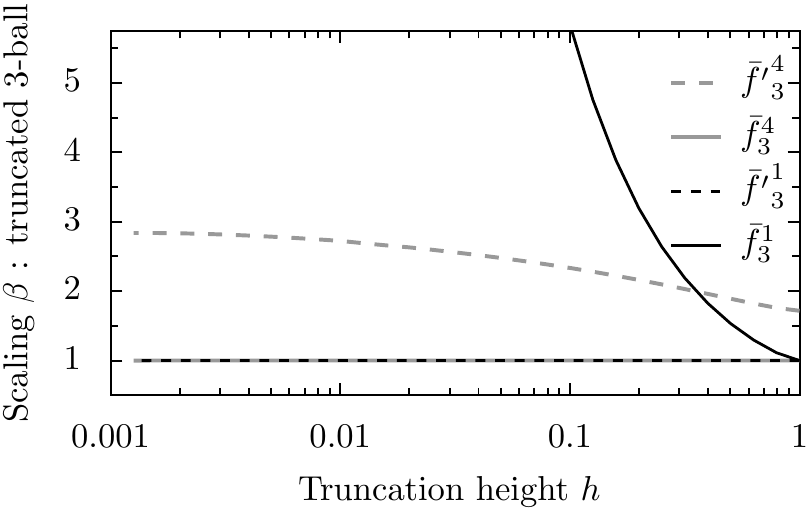} 
	\caption{ \label{truncated_pwlaw}
	Fitted power law $\beta$ for $\bar{f}_3^p(x)$ and $\bar{f'}_3^p(x)$ in a truncated 3-ball (see Fig. \ref{illus_shapes}) as a function of the truncation height $h$.
	}
\end{figure}

Because the scaling behaviour of $\bar{f}_n^p$ depends so strongly on the dimension $n$, it is interesting to consider what happens when one dimension becomes arbitrarily small. We thus integrated Eqs. \ref{def_dist},\ref{def_force} in a $3$-ball symmetrically truncated along its $Oy$ axis at a height $h$, see Fig. \ref{illus_shapes}. We found that the expected scaling law for $\bar{f}_3^{4}$ still held even for $h\rightarrow 0$, Fig. \ref{truncated_pwlaw}. Although this was expected from  $\bar{l}_n^n$ being space invariant in any convex space, it is interesting to see that a $(n)$-dimensional system does not behave as $(n-1)$-dimensional system when one dimension becomes infinitesimally small.

\begin{figure}[t]
	\includegraphics[width=\linewidth]{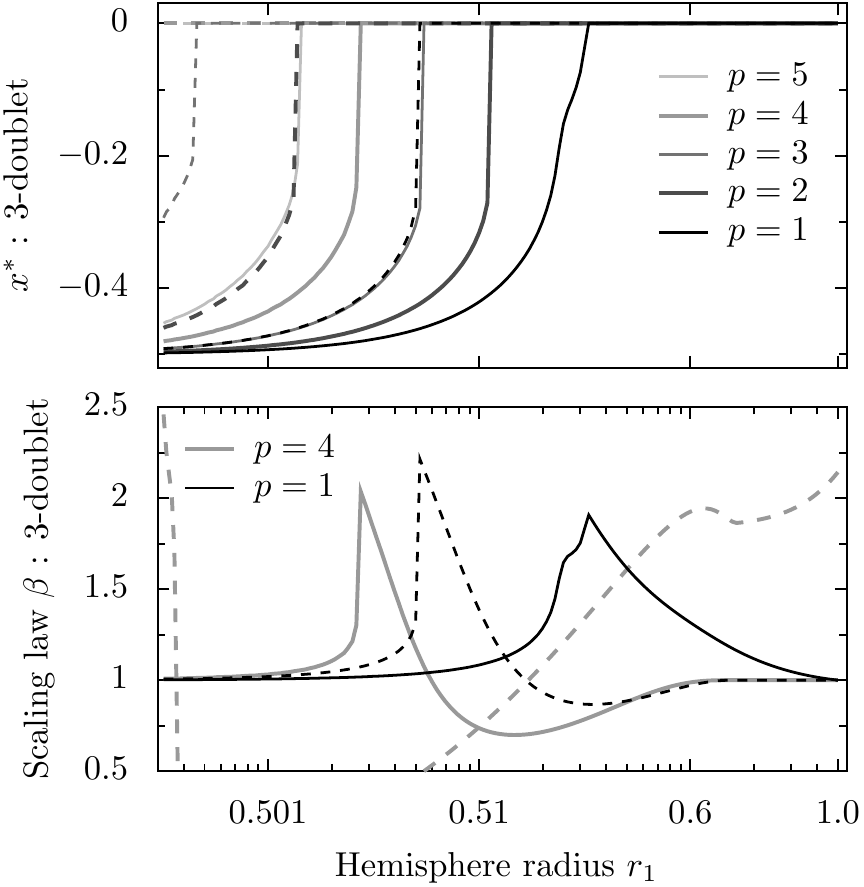} 
	\caption{ \label{doublet_pwlaw}
		{\em Top} : equilibrium position $x^*$ such that $\bar{f}_3^{p}(x^*)=0$ (solid lines) and  $\bar{f'}_3^{p}(x^*)=0$ (dashed lines) as a function of the hemisphere radius $r_1$ (log scale diverging at $r_1=0.5$). $r_1$ is defined such that $r_1=1$ for a single sphere and $r_1=0.5$ for a closed doublet. 	{\em Bottom} : Scaling law of  $\bar{f}_3^p(x)$ (solid lines) and  $\bar{f'}_3^p(x)$ (dashed lines) as a function of $r_1$.
		}
\end{figure}

\paragraph*{Centering in non-convex spaces}
Eventually, we considered a non convex shape by considering a doublet of spheres, truncated on their $Ox$ axis at $x=0$, mimicking the geometry of a dividing cell, see Fig. \ref{illus_shapes}. We took 3D spheres of radius $1/2 \le r_1 \le 1$, truncated at $x=0$ and with centers at $x=-1+r_1$ and $1-r_1$, see Fig. \ref{illus_shapes}. For a such shape, all the volume is not visible by any point in the sphere, and we do not expect $\bar{l}_n^n$ to be space invariant.  Indeed, we found that the violation of the scaling law initially increases as  $r_1$ decreases from 1, see Fig. \ref{doublet_pwlaw}, bottom. As the doublet closes ($r_1 \rightarrow 1/2$), the scaling laws are restored because the visible space tends towards a single sphere.

We would also expect the center to be at $x^*=0$ for $r_1=1$ (a spherical cell) and $x^*=\pm0.5$ for $r_1=0.5$ (a perfect doublet). Indeed, we find a transition of $x^*$ from $0$ to $0.5$ when $r_1$. This transition happens after a threshold value of $r_1$ that depends on $p$ ; whether this transition is continuous or not also depends on $p$, Fig. \ref{doublet_pwlaw}, top. This thus resembles a sub-critical pitchfork phase transition. Overall, the smaller $p$, or the larger $n$, the smaller $r_1^*$. This is true for both autonomous and surface-mediated centering.





\paragraph*{Discussion}
We first showed that maximizing the distance$^{p+1}$ to the surface is analog to pulling forces proportional to distance$^p$, allowing us to focus on pulling forces, that are known to take place in pronucleus centering. Pulling strategies always promotes centering, while maximizing distance$^p$ centers only if $p>n$.  We  found interesting scaling laws : $\bar{f}_n^{n+1}(x)$ and $\bar{f'}_n^1$ are linear in any convex space. A direct consequence of this result is that a cropped $n$-dimensional system does not behave as a $(n-1)$-dimensional system even if one dimension tends to zero. Interestingly, in the case of autonomous centering, hypo-elastic forces can yield a net hyper-elastic centering force.

Here we did not discuss pushing strategies with $p<0$. This usually yields a rather inefficient centering, with diverging forces at the boundary and small forces closer to the center. Moreover, they do not exhibit the rich phenomenology we observed with $p>0$.

These scaling laws are violated in non-convex shapes. For those, below a threshold of convexity, there exists a stable asymmetric state. This would hinder the centering of the aster in embryos if centering were to happen after the constriction of the cell into two daughter cells - which does not seem to happen as centering always occur before constriction. The most favorable autonomous centering force appears to be $\bar{f}_n^{n+1}$, being both elastic for any shape (hence allowing smooth centering) and efficient at finding the center except for extremely non-convex shapes, Fig. \ref{truncated_pwlaw}, top. Here we did not consider the possibility to maximize the {\em minimum} distance to the surface \cite{garcia2007poles}, notably because in the case of a non-convex shape such as a doublet, it will always find the center of the hemispheres rather than the center of the space.

In the case of the centering of an aster, centering could also be achieved in strongly non-convex shapes if each radial element (microtubule) generates other elements. This would allow the aster to explore the space non directly visible by the aster center. This is biologically possible as the filaments forming the aster (microtubules) can be used as templates for the nucleation of new filaments ; this could be significant during aster growth \cite{ishihara2016physical}. For autonomous agents using  sound or electromagnetic waves to measure distances, this might be achievable by using secondary reflections, albeit making the process more complex.

\paragraph*{Acknowledgements} The authors would like to thank Hirokazu Tanimoto for the valuable suggestions, François N\'{e}d\'{e}lec for suggestions on the manuscript, and Milan Lacassin for noticing the hypo-elasticity of $\bar{f}_3^3$, thus inspiring this work. S.D. was supported by a CNRS-Momentum fellowship and N.M by an ERC consolidator grant (Forecaster 647073).

\bibliographystyle{unsrt}
\bibliography{bibtex_centering}

\end{document}